# Quantum Scent Dynamics (QSD): A new composite model of physical particles

György Darvas[(1)] and Tamás F. Farkas[(2)]

(1) *Institute for Research Organization of the Hungarian Academy of Sciences*, and *Symmetrion*, 18 Nádor St., P.O.Box 994, Budapest, H-1245 Hungary E-mail: darvasg@iif.hu
(2) 46 Rózsa St., Budapest, H-1064 Hungary

*Abstract*: The paper introduces an alternative rishon model for a composite structure of quarks and leptons. The model builds matter from six basic blocks (and their antiparticles). For this reason it introduces new properties of rishons, called „scents", that can take two values, called masculine and feminine scents, which can appear in three colours both. The Quantum Scent Dynamics (QSD) model calculates new electric charges for the rishons. Then it discusses the construction of the known families of particles from scents, as well as the constraints and advantages of the proposed hypothetic model.

*Introduction*

It was about in 2001, when Tamás F. Farkas and I analysed one of his graphics. He designs sophisticated graphics with complex symmetries. I realised, that the given graphic could be used to model a baryon (e.g., a proton). It consisted of three units, like three quarks connected to each other (Figure 1, one can gaze at this wonderful graphics also at the cover design, as well as in pp. 396 and 507 of Darvas, 2007).

There seemed only one problem: the model was too symmetrical. Namely, the three „quarks" were identical, although of different colour. The common work started with this. Description of the model and some preliminary assumptions for possible consequences were published in (Darvas and Farkas, 2006). A physical analysis follows here.

First we decomposed the „baryon" unit into three composite units, called „quarks", and distorted the symmetry of the graphics of „quarks". We endowed them with „flavours", as well as distinguished two graphical units according to two different spin states. The flavour of the quarks was denoted by graphical signs, while the spin was denoted by concave and convex representation of the unit (Figure 2). As regarded the colours, for aesthetic reasons we applied yellow, instead of green of the common RGB colour system. Since colour is symbolic name of a property, RYB colour system is as much accepted in physics as RGB.

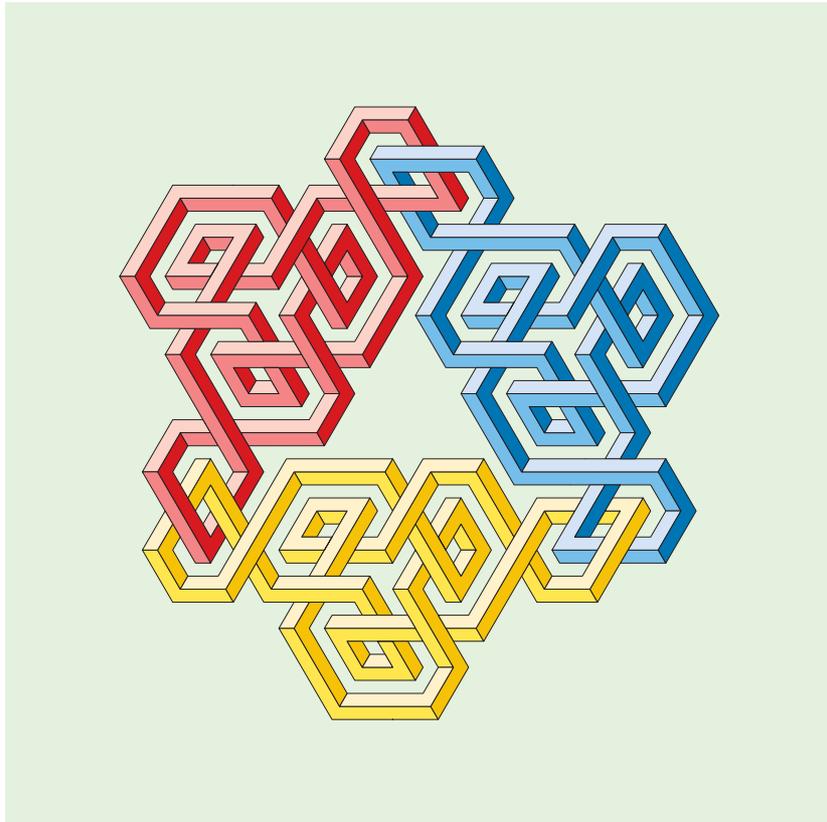

**Figure 1**

*Spin*

There are two quarks represented on Figure 2: spin up and spin down. A spin up quark is represented by a convex graphical unit. A spin down quark is represented by a concave graphical unit (like the opposite three neighbouring faces of a cube).

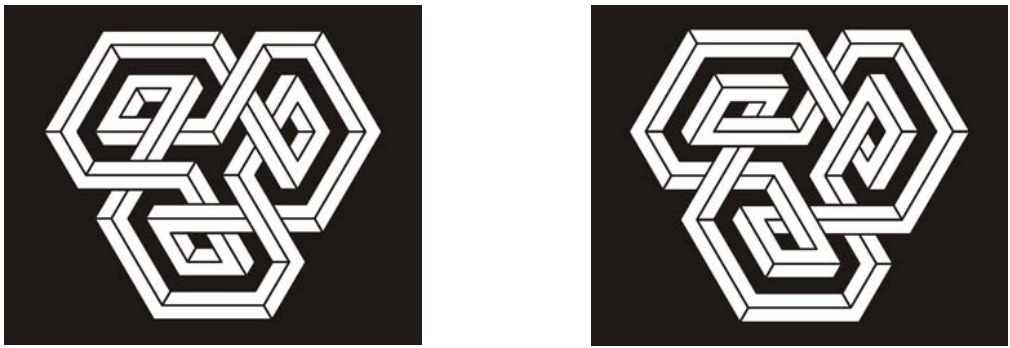

**Figure 2**
Graphical unit to represent a *quark*. *Spin up* (left) and *spin down* (right) quarks.

*Flavour*

To represent „flavour" we used the property of the „too symmetric" quark-model that a quark was graphically composed of three identical „sub-units". These sub-units contained a loop twist. These twists are all left-handed in Figure 1, as well as in the spin up unit in Figure 2, while they are right-handed in the representation of the spin down unit in Figure 2. The combinations of the left- and right-twists in a spin-up quark are represented in Figure 3.

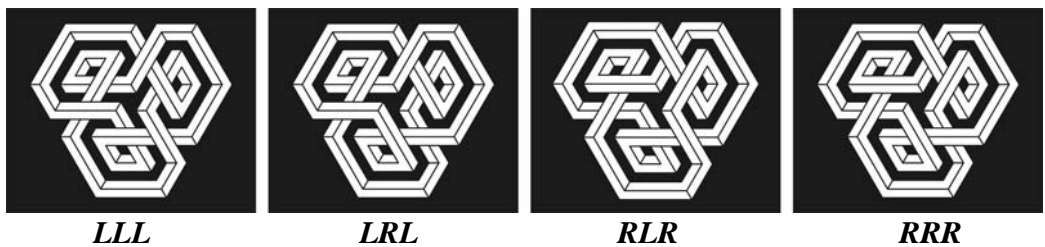

    ***LLL***        ***LRL***        ***RLR***        ***RRR***

**Figure 3**
The four possible combinations of *Left* and *Right* twists in a graphical unit.

All the three loop twists in the graphical unit of a quark can be left- (***L***) or right-handed (***R***). One unite can geometrically consist of 0, 1, 2 or 3 left-handed, and 3, 2, 1 or 0 right-handed twists, respectively. No other combinations can be distinguished from each other. All the four combinations of the twists may appear both in spin-up and spin-down units, respectively. For simplicity, later on we will deal only with one of the spin states of the units.

According to the Pauli principle the *LLL* and the *RRR* (the geometrically most symmetric) graphical units (Figure 3) cannot represent real physical particles. There remain two (*LRL* and *RLR*) physically possible graphical units. (Due to the rotational symmetry of the graphical unit, *LRL*, *LLR*, and *RLL* permutations are indistinguishable, i.e., identical.) According to our model, they represent hypothetic unbound quarks.

Then we built in the graphical model the phenomenon, that quarks never appear alone, only clipped to others, i.e., in bound states. According to our present knowledge and among natural (not-extreme) physical conditions nature produces only bound quarks. Always three quarks form a *baryon*, and always a quark and an anti-quark form a *meson*. The binding-clip to form a baryon can be joined to two of the three sub-units of the quark. (Shown in Figure 4 this clip has not too much concrete physical meaning, rather a graphic representation of binding.) As one can choose two of the three twists in three different ways, the number of possible combinations triples. Any of the graphical units can be clipped to their two neighbours through three different permutations of the clips on their sub-units.

By the variation of the two allowed twist combinations with the three permutations of the possible clips, (having rejected the graphical units excluded by the Pauli principle), we could represent six different flavour quarks (cf., Table 1). Let's agree in a convention to make the following correspondence between the individual variations and flavours, based on a classification of the placement of clips (marked by "-"):

|  | (flavour) |  | (flavour) |
|---|---|---|---|
| *-LRL-* | = *u* (up) | *-RLR-* | = *d* (down) |
| *-LLR-* | = *c* (charm) | *-RRL-* | = *s* (strange) |
| *-RLL-* | = *t* (top/truth) | *-LRR-* | = *b* (bottom/beauty) |

**Table 1**

When a rotational symmetric quark (or the graphical unit representing it) enters into a binding with two others to form a baryon, it loses its certain symmetries. Note, this is a more general geometric principle of nature, valid not only for the quarks: a single object has more symmetry properties, than as a part of a system. (However, there may appear new symmetry properties in the more complex systems; Darvas, 2005.) When making correspondence between the individual flavours and bind permutations let's assume, that nature prefers their symmetric joining (i.e., *-LRL-* and *-RLR-*), this being the reason for the most frequent appearance of the *u* and *d* quarks in the usual matter.

*Colour*

All flavours of quarks can appear in three different forms. They can be represented by the symbolic "colours", that physicists used to call them. Figure 4 represents the binding of the different flavour units in one of the possible three colours each. We represent the six colour-triplets and the six flavours in the following way:

| (yellow) | (red) | (blue) | (flavour) |
|---|---|---|---|
| *-LRL-* | *-LRL-* | *-LRL-* | = *u* (up) |
| *-LLR-* | *-LLR-* | *-LLR-* | = *c* (charm) |
| *-RLL-* | *-RLL-* | *-RLL-* | = *t* (top or truth) |

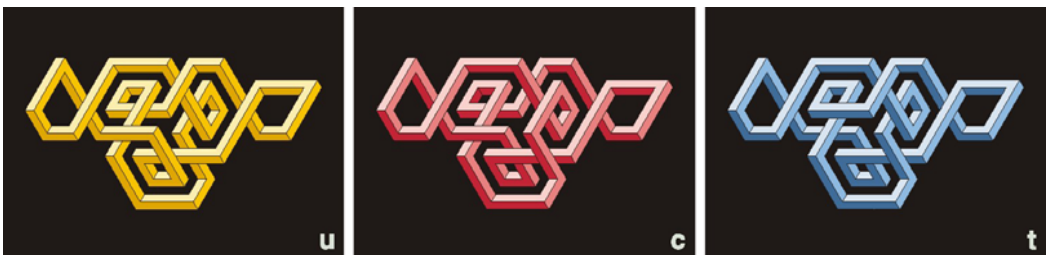

The 3 possible pairs of clips attached to the **LRL** quarks
(each represented in one colour only)

|   (yellow)   |   (red)   |   (blue)   |   (flavour)   |
|---|---|---|---|
| -*RLR*- | -*RLR*- | -*RLR*- | = ***d*** (down) |
| -*RRL*- | -*RRL*- | -*RRL*- | = ***s*** (strange) |
| -*LRR*- | -*LRR*- | -*LRR*- | = ***b*** (bottom/beauty) |

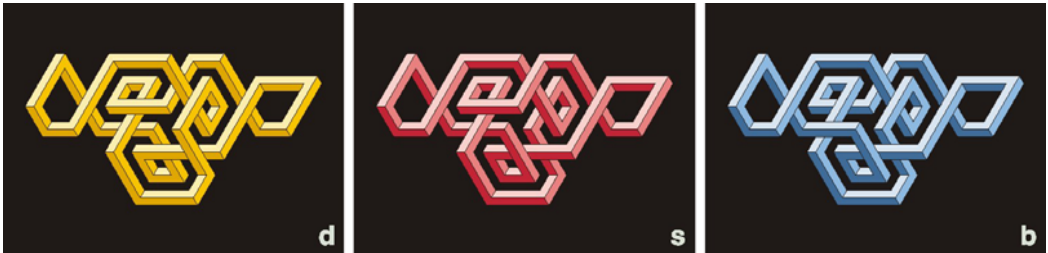

The 3 possible pairs of clips attached to the ***RLR*** quarks
(each represented in one colour only)

**Figure 4**

For simplicity in representation we will graphically represent the antiparticles of the individual particles by their anticolours. (We define anticolours in the following way: if a given colour is represented by the values (r, y, b) where r, y, b ∈ [0, 255], its anticolour is characterised by (255-r, 255-y, 255-b) ).

*Bound quarks*

E.g., the *u* quark is represented by a graphical sub-unit, where the two bounding clips join to two left-handed loop twists, and there is a right-handed loop twist in it, which is not endowed by a clip; the *s* quark has one clip at a right-handed loop twist, another clip at a left-handed loop-twist, and there is a further right-handed twist, which is not endowed by a clip, and so on.

*Scent*

The above introduced model represented the known properties of the physical particles used in the Standard Model. However, the reader could recognise, that we used such auxiliary terms, like "twist" and "sub-unit", which did not correspond to any concept in the Standard Model (SM) of physics. In the following we give meaning to these terms. Although, they do not correspond to any experimental fact, they stand not without example in the history of modern physics. The novelty is, that certain old assumed properties get a new meaning and value in the new model proposed by us.

To remain in the field of senses (in other used words, in the field of quantum aesthetodynamics) in naming the properties, we call the property of the sub-units marked by the possible two opposite directions of twists *scent*. We call the two values that scent can take as "masculine" scent (according to a proposed convention for the Left-handed direction of the twist, Figure 5), and as "feminine" scent (according to a proposed convention for the Right-handed direction of the twist, Figure 6). In the following we will use this newly introduced terminology. Instead of Left-handed and Right-handed twists, we will speak of *masculine* and *feminine scents*.

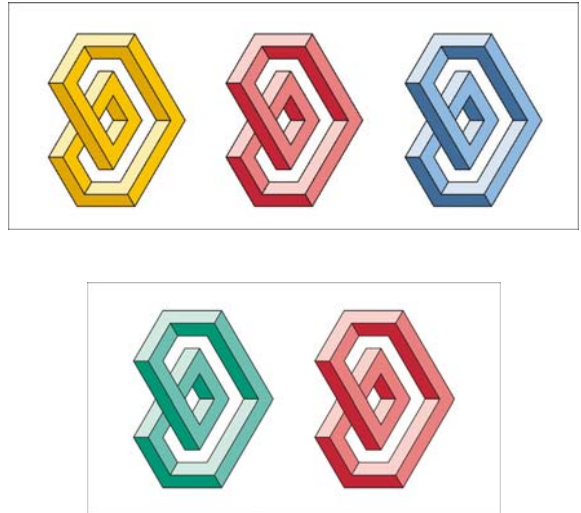

**Figure 5**
Upper row: The *L* (masculine) rishons in all the 3 colours
Lower row: An *L* (masculine) anti-rhishon and rishon

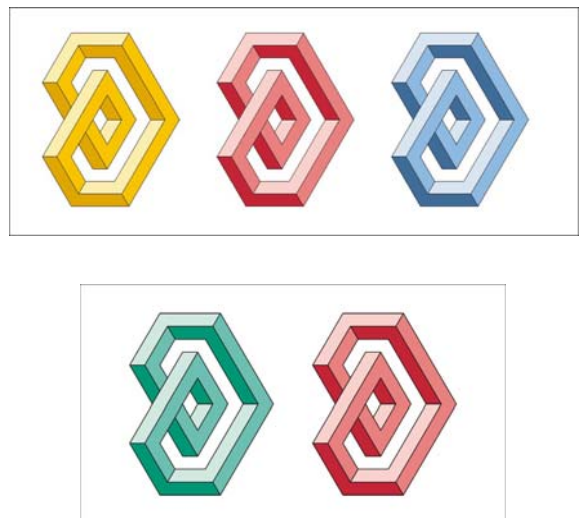

**Figure 6**
Upper row: The *R* (feminine) rishons in all the 3 colours
Lower row: An *R* (feminine) anti-rhishon and rishon

*Rishons*

We identify the "sub-units", that constitute a quark in the model, with *rishon*s. Rishons are hypothetic particles, introduced by H. Harari (1979). However, the rishons introduced by our QSD model differ in their properties from those called by Harari as "Tohu" and "Vohu", and he assumed them to have electric charges (-1/3) and 0 (and his anti-rishons would have the charges 0 and 1/3). (Earlier aborted attempts called hypothetic composite blocks of matter as preons (or prequarks, Pati and Salam), subquarks (Terazawa, Chikashige and Akama), maons (Glashow), alphons (Ne'eman) and quinks (t'Hooft). A model similar to Harari's rishons by M.A. Shupe, 1979, worth also mentioning.)

We draw a similar model. However our model paints a bit different picture on the rishons, than Harari's one. The quarks in our QSD model are composed of a triplet from among the two rishons (Figures 3-4), and leptons are composed of a pair of a rishon and an anti-rishon, represented in Figures 5-6. Since they differ from Harari's rishons, we will denominate them with different name. According to their scent, we call them *masculine* and *feminine rishons*. As the graphical model suggests, masculine and feminine rishons are not identical with Tohu and Vohu. They are different – actually – hypothetical particles.

All the represented quarks can be composed of the two graphical units in Figures 5 and 6. Masculine scent is represented by a *L*eft-handed twisted rishon, feminine scent is represented by a *R*ight-handed twisted rishon. Thus, for the sake of easier understanding, we will continue to mark masculine scent by letter *L*, and feminine scent by letter *R* – at least in this paper. Later we propose to use the letters *M* and *F*.

Similar to Harari's model we want also to compose the electric charges of the quarks to be (2/3 for *u*, *c*, *t*) and (-1/3 for *d*, *s*, *b*), as they are set in the SM. The *u*, *c*, and *t* quarks (charge +2/3) are composed of two *L*eft-handed twists and a *R*ight-handed twist (two masculine and a feminine scent) rishons. The *d*, *s* and *b* quarks (charge -1/3) are composed of two *R*ight-handed twists and a *L*eft-handed twist, or in other words, two feminine and a masculine scent rishons (Figure 4).

This combination demands, that the electric charge of the masculine scent rishon be 5/9, and the electric charge of the feminine-scent rishon be -4/9.

Observe that these graphical units are asymmetric, each. Their peculiarity is, that one can compose visually symmetric units (quarks, see Figure 2) from three, individually asymmetric components (rishons). This asymmetry predicted, that one could not expect symmetric properties devoted to these primary components (cf., their electric charges +5/9 and -4/9 respectively).

Anti-rishons have the opposite sign charges of the rishons, and will be represented by the complementary colours of the respective rishons (cf., Figures 5 and 6, lower rows).

We saw above, how could we represent the families of baryons built up of quarks consisting of rishons. The rishon model read from Farkas' graphics – similar to Harari's model – is suitable to represent all types of particles. Now we have a look at the different families of particles separately.

*Leptons*

Let's represent the lepton members of the electron generation by red and anti-red (its complementary colour, i.e., cyan) colours.

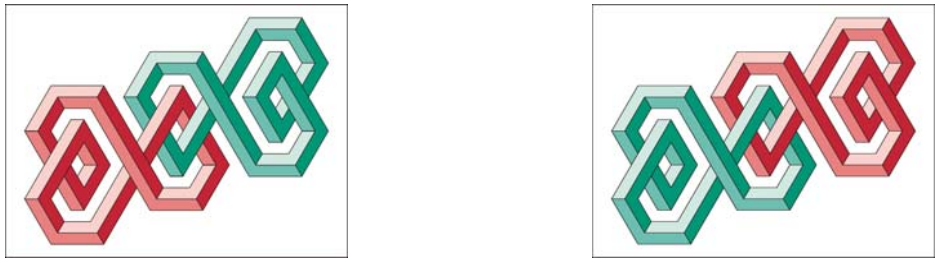

**Figure 7**

The electron ($e^-$)          The positron ($e^+$)

An electron (Fig. 7 left) can be composed of a feminine and an anti-masculine scent rishon ($\bar{L}R$). Its charge is -1 [=(-5/9) + (-4/9)]

A positron (Fig. 7 right) can be made of a masculine and an anti-feminine scent rishon ($R\bar{L}$). Its charge is +1 [=4/9 + 5/9]

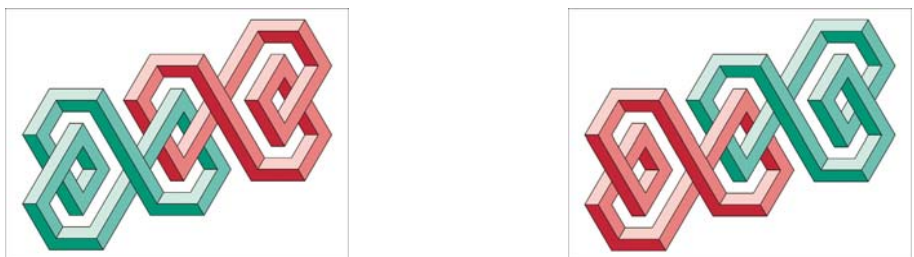

**Figure 8**

The neutrino ($\nu$)          The anti-neutrino ($\bar{\nu}$)

A neutrino (Fig. 8 left) can be composed of a feminine and anti-feminine scent rishon pair ($R\bar{R}$). Its charge is 0.

Similarly, an anti-neutrino (Fig. 8 right) can be made of a masculine and anti-masculine rishon pair ($L\bar{L}$). Its charge is also 0.

This method properly represents the leptons colourless. We propose, that the *muon* and *tau* lepton generations could be represented by the other two colours (yellow and blue) and their complementary colours.

*Spin of the rishons*

For the sake of consistancy of the QSD model we cannot avoid to attribute a spin to the single rishons. We have no reason to not presume them to behave as fermions.

Rishons depicted in Figures 5 and 6 are all concave. According to a convention introduced in Figure 2, they depict spin down states (cf., right image in Figure 2). Convex, i.e., spin up state rishons can be represented in a similar way. If the spin of rishons can occupy the $J = +(1/2)$ and $J = -(1/2)$ states, quarks can be made of two up and a down, or an up and two down spin rishons to ensure the half-spin states of the quarks. These slightly distort aesthetic symmetry, but can be represented in our graphical model.

There poses more problems for the model to ensure half-spin states for the *leptons*, which consist of rishon-antirishon doublets. One cannot produce half-spin resulting states composed of two half-spin units without further consideration. Let we consider the following assumption: the bounding clips joining two rishons own an additional $J = \pm(1/4)$ absolute value spin, which is parallel with the spin of the given rishon, and antiparallel with the spin of the given antirishon. (In the graphical representation the clips can be depicted as concave or convex independent of the respective rishons. There is no explanation, why should we assume such a correlation between the electric charge and the spin.)

*Leptons* consist of *antiparallel spin rishon-antirishon pairs* (Table 2). In this case the resulting spins of the leptons will take a $J = \pm(1/2)$ value, as expected. Note, that the model is unable to predict and explain, why do neutrinos and antineutrinos appear only in one spin state, and never observed in the opposite.

*Parallel spin rishon-antirishon pairs* compose *bosons* (see below, Table 3).

If the binding clips of the rishons own an additional spin in leptons, we must assume the same for their binding clips in the triplets, forming the quarks. However, quarks have two binding clips, and further, the construction provided by the model for composing the quarks ensures that the two binding clips join to two opposite spin rishons – without introducing any further restriciton in the model – so the additional spins of the even number binding clips are antiparallel and neutralise each other.

**Table 2**:
Possible spin compositions of leptons on the example of the electron generation. Leptons are built of antiparallel spin rishon and antirishon pairs. (Note, that the model is unable to exclude neutrinos and antineutrinos in both spin states.) Leptons of the muon and tau generations can be represented by the rest two colours.

(For spin considerations, *graviton* should consist of four scents, e.g., $L\bar{L}R\bar{R}$ all the four in spin up position.)

*Quarks*

Figure 4 depicts all the six flavour quarks (each in one colour only) in their bound representation. Figure 9 presents the *u* and the *d* quarks in their – hypothetic – unbound state in all the three colours both.

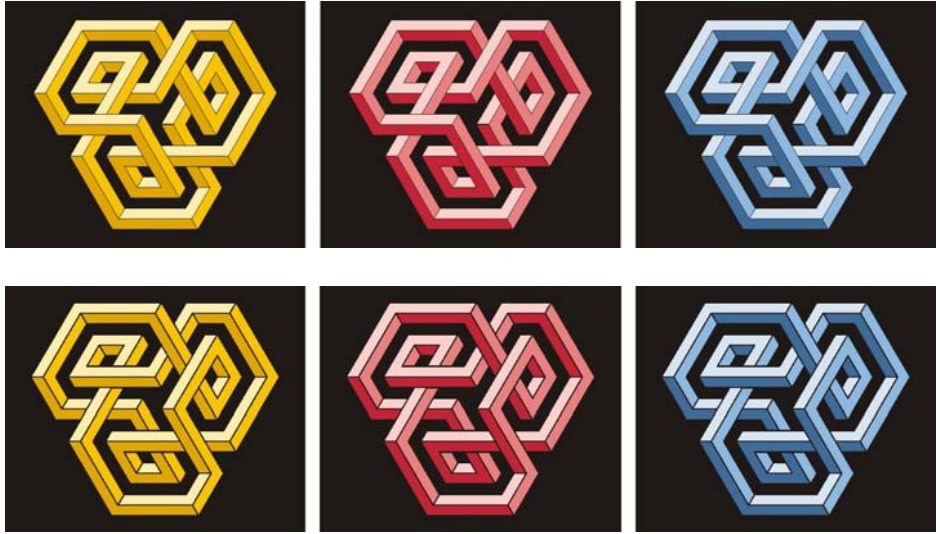

**Figure 9**
Single quarks in – hypothetic – unbound state
Upper row: The *u* (*LRL)* quark in all the 3 colours.
Lower row: The *d (RLR)* quark in all the 3 colours

*Baryons*

We saw above, how can the different flavours be represented in our model by the variation of binding clips to the individual scents. In this way, we identified the six flavours, that may appear in three colours each. *Baryons* consist of three quarks. Binding the different flavour quarks to each other in all the possible ways, we can produce the complete family of baryons. Figure 10 represents the proton (*udu*) and Figure 11 the neutron (*dud*).

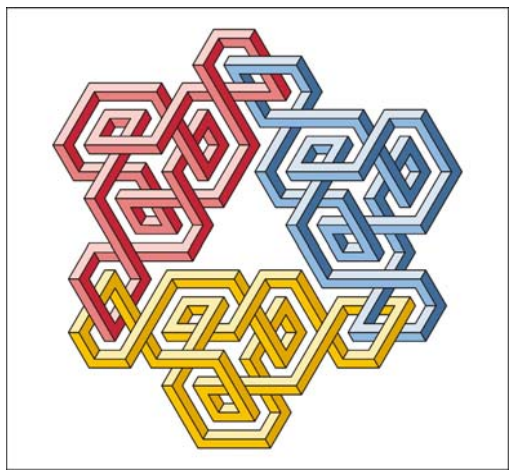

**Figure 10** The proton (*udu)*

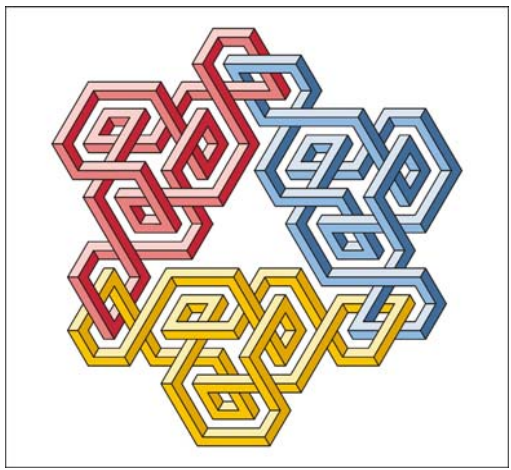

**Figure 11** The neutron (*dud)*

The *u* and the *d* quarks within the model of the proton and the neutron can be checked by identifying them at the Figure 4 (left). (We apologise that Figures 10 and 11 do not make a distinction between the up and down spin quarks, however, it is possible, as we showed above.)

According to the SM of physics, the individual quarks permanently change their colours, so that at any given moment, all of them appear in different colours. This process can be observed on the model of the proton and the neutron in an animated form at the sites http://members.iif.hu/visontay/ponticulus/images/darvas/proton_anim.gif and http://members.iif.hu/visontay/ponticulus/images/darvas/neutron_anim.gif respectively. as well as at *http://us.geocities.com/symmetrion/*QCD/proton123.swf and *http://us.geocities.com/symmetrion/QCD/neuton123.swf* .

*Mesons*

*Mesons* are pairs of a quark and an anti-quark. Anti-quarks – similar to the anti-rishons – are represented in this (QSD) model by their complementary colours. In accordance with Quantum Chromodynamics (QCD) anti-quarks are represented (graphically) identical with their corresponding quarks, however, their colour is the complementary colour of the respective quark. (As mentioned above, the numerical hexadecimal inverse of a code of a colour determines an inverse-code colour, and this is called the complementary colour of the original.) When forming a meson, quarks are bound only to one other, thus – in our model, for symmetry considerations – they join to the anti-quarks next to that scent, which does not take part in binding into a baryon. (E.g., a *u* quark joins to two quarks next to its two *L*eft-handed twists, and to an anti-quark next to its *R*ight-handed twist.) We can observe an example for modeling two mesons in Figure 12.

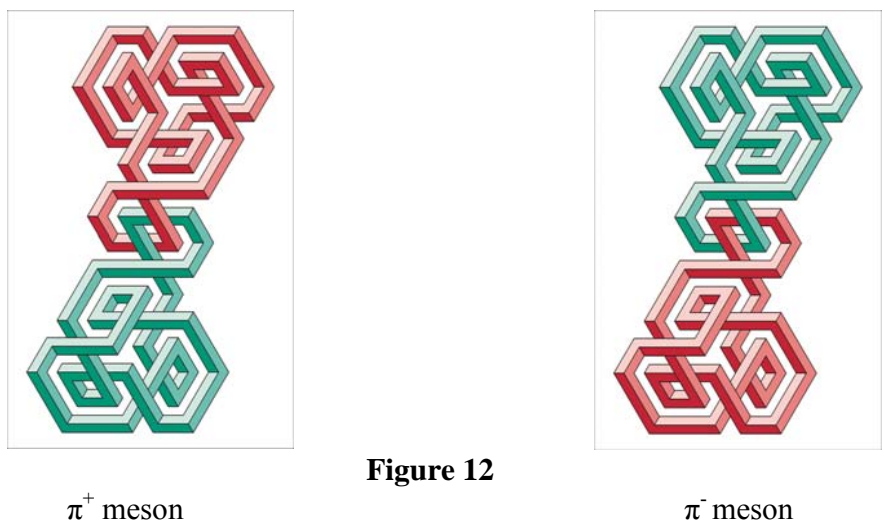

**Figure 12**

π⁺ meson            π⁻ meson

The π⁺ meson (Figure 12 left) consists of a *u* (here red) and an anti-*d* ($\bar{d}$) quark (here cyan). A *u* quark is coupling with an antiquark to form a meson joining through its *R*ight- handed twist, while a *d* or $\bar{d}$ quark through its *L*eft- handed twist. The charge of a π⁺ meson equals to +1 (= 2/3 + 1/3).

Let us compose the model of the π⁻ meson. The π⁻ meson (Figure 12 right) consists of an anti-*u* ($\bar{u}$) quark and a *d* quark. If the *d* quark takes the colour red, the anti-*u* ($\bar{u}$) quark takes the complementary colour, that we mark with cyan. The charge of a π⁻ meson equals to -1 [= (-1/3) + (-2/3)].

*Intermediate weak bosons*

Substituting quarks and leptons by their composite rishon triplets and doublets in weak interactions, we can observe that charged weak current bosons *W* flip one of the scents in each reaction to the opposite, keeping the colour of the respective rishons or antirishons intact. To apply the model for weak reactions, $W^+$ can be represented as parallel spin $\overline{R}L$, that may flip a feminine scent rishon *R* into a masculin scent rishon *L*, or a masculine scent antirishon $\overline{L}$ into a feminine scent antirishon $\overline{R}$. Its resulting spin is 1 integer and electric charge is 1. In a similar way, $W^-$ can be represented as parallel spin $\overline{L}R$, that may flip a masculine scent rishon *L* into a feminine scent rishon *R*, or a feminine scent antirishon $\overline{R}$ into a masculine scent antirishon $\overline{L}$. Its resulting spin is 1 integer and electric charge is -1 (Table 3).

Weak current bosons can appear in all the three colours, according to the actual colour of their composite rishons and antirishons. We do not know, what selection rule of the nature restricts the combination of a rishon with a different colour opposite scent antirishon to form a boson, we know only, that such a combination would change the colour of the rishon affected by that boson; this would demand a simultaneous colour change of the other two rishons in the same quark, and that would involve the demand for the change of colour of another quark within the hadron, what chain of transformations can hardly be imagined. Thus, only identical colour rishons and antirishons can compose a weak current boson.

Examples for weak interactions can be described by the QSD model in following ways:

$$W^+ \to e^+ + \underline{\nu}_e \qquad \overline{RL} \to \overline{RL} + R\overline{R}$$
$$W^- \to e^- + \nu_e \qquad \overline{LR} \to \overline{LR} + L\overline{L}$$

Note, that the $\overline{RL}$ and $\overline{LR}$ particles of the left and the right sides of the arrows in the above reactions correspond to different particles. There are bosons at the left side, and fermions on the right. During the reaction there happened a spin change, according to the upper rows of the Tables 2 and 3. While antiparallel spin rishon-antirishon pairs composed leptons, their parallel spin pairs compose bosons.

One can describe further reactions in a similar way. E.g., reactions in $p\overline{p}$ collisions can be described as follows:

$$u + \overline{d} \to W^+ \qquad LRL + \overline{RLR} \to L\overline{R} + R\overline{L} + L\overline{R} = 2W^+ + W^- = W^+ \quad \text{or}$$
$$u + \overline{d} \to W^+ \qquad LRL + \overline{RLR} \to L\overline{L} + R\overline{R} + L\overline{R} = 2\gamma + W^+$$

$$d + \overline{u} \to W^- \qquad RLR + \overline{LRL} \to R\overline{L} + L\overline{R} + R\overline{L} = 2W^- + W^+ = W^- \quad \text{or}$$
$$d + \overline{u} \to W^- \qquad RLR + \overline{LRL} \to L\overline{L} + R\overline{R} + R\overline{L} = 2\gamma + W^-$$

These reactions provide a possibility to check the model. Although we detect (the decayed traces of) a single $W$ resulted in the reaction, it is possible to find trace of the short lived excitations of the $W^+ W^-$ pair, that annihilate each other immediately.

Let we see the two steps reaction of a muonic neutrino and a neutron with the mediation of the emission of a $W^+$:

$$\nu_\mu \to W^+ + \mu^+ \quad \to \quad W^+ + n \to p \quad : \qquad R\overline{R} \to \overline{RL} + \overline{LR} \quad \to \quad \overline{RL} + RLR \to LRL$$

A $\beta$ decay can be described in two steps, mentioning only the affected quarks:

$$d \to W^- + u \quad \to \quad W^- \to e^- + \overline{\nu} \quad : \qquad RLR \to \overline{LR} + LRL \quad \to \quad \overline{LR} \to \overline{LR} + L\overline{L}$$

Note in the above reaction again the spin changes within the $\overline{LR}$ structure. The model can explain also, why may $W$ decay either into two quarks+… or to two leptons+… .

According to the bottom row of the Table 3, the bosonic (integer spin) states of both the feminine and the masculine rishon-antirishon pairs ($R\overline{R}$ and $L\overline{L}$) correspond to the neutral weak current boson. Keeping the rules established in the model, both pairs should describe $Z^0$. Similar to their leptonic brothers, we will consider them the antiparticles of each other and mark them with $Z^0$ and $\overline{Z}^0$. Their reactions justify this assumption, since they appear both in $R\overline{R}$ and $L\overline{L}$ forms.

Nevertheless, the $R\overline{R}$ and $L\overline{L}$ rishon-antirishon pairs seem the most misterious particles in the model. As we will see, they play role in more constructions than the others. The mistery started at their neutrino combination, which does appear only in one of the possible spin positions. We have no reason to presume, that their bosonic forms could not appear in all the possible integer spin combinations. In the leptonic combinations a single colour neutrino-anineutrino pair belongs to one of the three individual generations. The bosonic forms of the same rishon pairs appear in all the three colours (and their combinations), and can take part in different reactions (see the $Z^0$ decays later). And their mystery does not end with this.

Examples for the role of the neutral weak current boson:

$u + \overline{u} \to Z^0$    the same reaction in our model:

$LRL + \overline{LRL} \to L\overline{L} + R\overline{R} + L\overline{L} = 2\overline{Z}^0 + Z^0 = \overline{Z}^0$

This reaction justifies our assumption on the distinction between $Z^0$ and $\overline{Z}^0$ (Table 3).

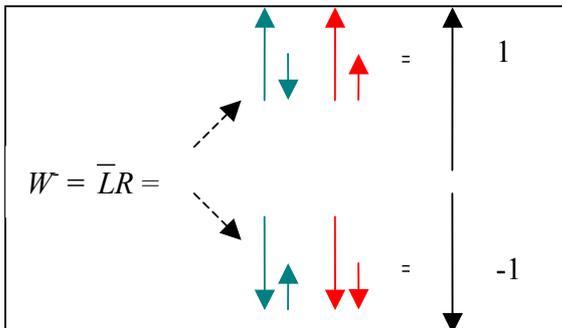
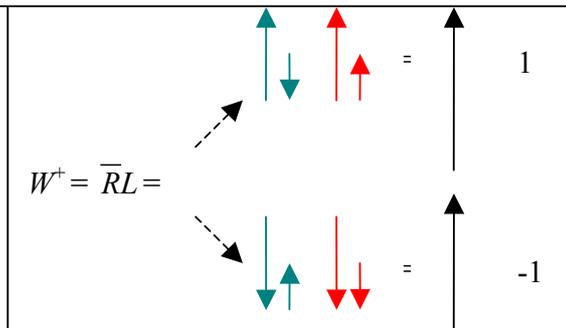
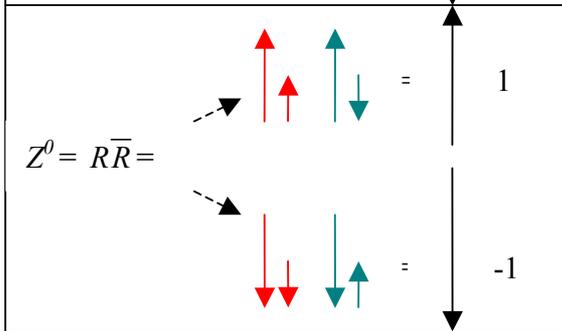
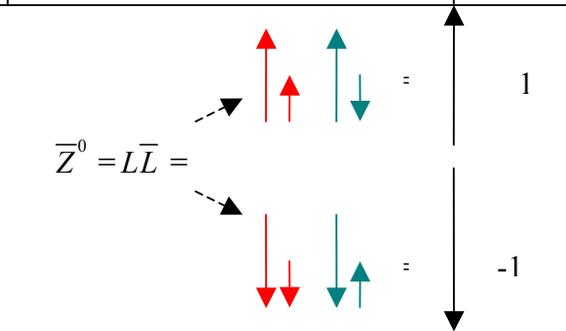

**Table 3**
Possible spin compositions of intermediate bosons. Bosons are built of parallel spin rishon and antirishon pairs.

The above seen $u + \bar{u}$ reaction in $p\bar{p}$ collisions can take place in another issue too:

$$u + \bar{u} \rightarrow W^+ + W^-$$

This reaction was expected in experiments, although it would violate spin conservation. Our model shows, that the above learned $L\bar{L}$ rishon-antirishon pair should appear in the reaction too.

$$LRL + \overline{LRL} \rightarrow L\bar{R} + R\bar{L} + L\bar{L}$$

The third boson annihilates as a massive particle and irradiates. This is the first sign, that $R\bar{R}$ and $L\bar{L}$ rishon-antirishon pairs in their bosonic form can appear not only in the form of the massive $Z^0$, but depending on the energy level in which the reaction takes place, they can represent $\gamma$ as well.

Investigating further the $Z^0$ mystery, we can observe, that $Z^0$ can decay all for an electron-positron pair, and muon-antimuon, tau-antitau pairs. Since the resulting pairs – at least according to our model – differ only in their colours, we can assume, that the outcome of the reaction depends on the actual colour of the rishon in the decaying $Z^0$ boson.

$$Z^0 \rightarrow e^- + e^+ \qquad Z^0 \rightarrow \mu^- + \mu^+ \qquad Z^0 \rightarrow \tau^- + \tau^+$$

Nevertheless, we can compare this with a fourth reaction:

$$\gamma \rightleftharpoons e^- + e^+ \qquad (= \bar{R}R + \bar{L}L = \bar{R}L + \bar{L}R)$$

Seemingly $\gamma$ can here also play a similar role, that $Z^0$. There depends on the energy level, which actualizes. And there arise the following two questions: which of the two pairs, $R\bar{R}$ or $L\bar{L}$ represents $\gamma$; as well as what makes a distincion in our model between $\gamma$ and $Z^0$? We can give a common answer to both questions. Unlike to $Z^0$, there is no reason to attribute any colour and any scent to $\gamma$; therefore $\gamma$ should be represented by a colourless and scentless combination of the $R\bar{R}$ and $L\bar{L}$ rishon-antirishon pairs. For it is massless, this combination will not cause a problem in the model. However, similar to several other vertices of the model, there is no explanation for the difference between the massiveness of the $Z^0$ and of massless state of the $\gamma$, both built of the same rishons. So cannot we explain, why is produced in one case a stable consturction ($\gamma$), while a short lived excitation ($Z^0$) in the other.

Let's investigate again a reaction of the above discussed muonic neutrino. Its collision now with a proton will be mediated by a $Z^0$ boson, and the reaction can be described in two steps again. The neutrino leaves unchanged, while the proton disintegrates to several hadrons.

$$\nu_\mu \to Z^0 + \nu_\mu \quad \to \quad Z^0 + uud \Rightarrow (\textit{different hadrons})$$

$$R\overline{R} \to R\overline{R} + \overline{R}R \quad \to \quad R\overline{R} + LRL + LRL + RLR \Rightarrow (\textit{different hadrons})$$

Note, in the left reaction the muonic neutrino disintegrates to a $Z^0$ and another muonic neutrino, while changing spins of the feminine rishons. In the second part of the reaction the $Z^0$ boson disintegrates, so that its feminine antirishon can flip the scent of different rishons all in the two *u* and the *d* quark of the proton, so allowing different products of the reaction. Different hadrons can be produced according to which of the *R*-s does it flip, and thus how does it permute the binding clips within the effected quark, i.e., how does it change those flavour.

*Gluons*

The role of gluons in our QSD model is the same like in the QCD of the SM. They change the colour of the rishons presumably in the same way like of the quarks in the QCD. The requirement to change the identical colour of the three rishons within a quark simultaneously is added to the old mystery, that governs to change the colour of the three quarks within a baryon simultaneously. We assume, that in strong interactions gluons act only on composite triplet states of rishons.

In the text we will mark the colour of the individual rishons by bottom indices, *b* for blue, *r* for red and *y* for yellow. The role of gluons is to change the colour of a respective particle. "Basic gluons" must change the colour of the rishons, flip their spin, but keep intact their scent and electric charge. This requirement involves to have two groups of *basic gluons* for the two scents. Therefore basic gluons can be represented in the bosonic form $G^L_{i,\bar{j}} = L_i \overline{L}_j$ (*i,j = b, r, y*) and $G^R_{i,\bar{j}} = R_i \overline{R}_j$ (*i,j = b, r, y*). They are electrically neutral, massless and have an integer spin. This algorythm produces 6-6 colour-changing basic gluons. We have to add to the set a colour-preserving basic gluon composed of a combination of the three (masculine) $L_i \overline{L}_j$ (*i,j = b, r, y*) and the three (feminine) $R_i \overline{R}_j$ (*i,j = b, r, y*) bosons, as well as their opposite spin counterpairs, to make the unchanged colour rishon flip back in its original state. Thus we get two groups of 8-8 basic gluons.

We can observe, that there appear the same rishon-antirishon pairs as basic gluons, which appeared in the $Z^0$ boson. However, $Z^0$ did not change the colour of the rishons affected by it in reactions, otherwise it could not affect only a single rishon within a quark, left the two others intact. (The three rishons within a quark must be left in the same colour during a weak reaction mediated by $Z^0$.) Therefore we must add to the explanations given to Table 3, that scents in $\overline{Z}^0$ must appear always in the coinciding colour combination $L_i \overline{L}_i$ (*i = b, r, y*) and in $Z^0$ as a coinciding colour combination of $R_i \overline{R}_i$ (*i = b, r, y*). This latter addition can give an explanation for the three respective decay modes of $Z^0$ into electron-positron, muon and tau pairs.

At this point we must return to that property of the QSD model, that leptons appear in one colour in each of the particle generations. If so, we must assume, that basic gluons should take part in all reactions in which a lepton interacts with particle members of another generation (e.g., when a muon decays into an electron, an electron neutrino and a muonic neutrino, etc.) This does not fit in the traditional model, where gluons were confined and interacted only strongly. Moreover, if this assumption holds, one must catch unconfined single gluons in these reactions. This may be another control of the model.

There is also no explanation, why are $L_i \overline{L}_j$ and $R_i \overline{R}_j$ massive $Z^0$ bosons if $i = j$, and massless gluons if $i \neq j$. Anyway, this is no less mysterious, than the old known decay of a massless $\gamma$ into massive electron-positron pair, or – to remain at the gluons – how can freed massless gluons disintegrate in a short run to jets of massive hadrons.

We have no clue for what selection rule limits the combination of different colour rishons into a triplet (to form a quark). We could assume, that – according to our presupposition (cf., above the composition of leptons) – different colour single rishons belong to different particle generations. However, then we must face the problem, why their all colour triplets take equal part in the formation of all flavours, irrespective of which generation do those flavours belong. One of the secret mysteries replaced by another. Choose, which is more sympathetic!

For the above introduced basic gluons must change the colour of the three rishons within a quark simultaneously, our model presupposes that the basic gluons can appear only in gluon triplets. These triplets correspond to the gluons identified in QCD. (Like scents in the respective quarks – cf. Table 1 – there appear two identical basic gluons in each gluon triplet in opposite spin position.) They must form two groups according to the number of feminine and masculine rishons in the individual quarks on which they act. The first (upper) group acts on the upper (Q = 2/3 charged) quarks (*u*, *c* and *t*), while the second (bottom) group acts on the bottom (Q = -1/3 charged) quarks (*d*, *s* and *b*). The two colour-preserving gluons are – in both groups – triplets composed of the respective combination of (two masculine and one feminine, as well as two feminine and one masculine, respectively) colour-preserving basic rishons with opposite spins (Table 4).

| Upper group of gluons | Bottom group of gluons |
|---|---|
| $(L_b\overline{L_r} + R_b\overline{R_r} + L_b\overline{L_r})\ (L_r\overline{L_b} + R_r\overline{R_b} + L_r\overline{L_b})$ | $(R_b\overline{R_r} + L_b\overline{L_r} + R_b\overline{R_r})\ (R_r\overline{R_b} + L_r\overline{L_b} + R_r\overline{R_b})$ |
| $(L_r\overline{L_y} + R_r\overline{R_y} + L_r\overline{L_y})\ (L_y\overline{L_r} + R_y\overline{R_r} + L_y\overline{L_r})$ | $(R_r\overline{R_y} + L_r\overline{L_y} + R_r\overline{R_y})\ (R_y\overline{R_r} + L_y\overline{L_r} + R_y\overline{R_r})$ |
| $(L_y\overline{L_b} + R_y\overline{R_b} + L_y\overline{L_b})\ (L_b\overline{L_y} + R_b\overline{R_y} + L_b\overline{L_y})$ | $(R_y\overline{R_b} + L_y\overline{L_b} + R_y\overline{R_b})\ (R_b\overline{R_y} + L_b\overline{L_y} + R_b\overline{R_y})$ |
| + 2 colour-preserving rishons (combinations of 2 masculine and 1 feminine basic neutral rishons both ) | + 2 colour-preserving rishons (combinations of 2 feminine and 1 masculine basic neutral rishons both) |

**Table 4**: The two groups of gluons

*Summary of the model*

*Rishons:* there are assumed a masculine-scent (left-handed) and a feminine-scent (right-handed) rishon.

*Scent*: chiral porperty, that distinguishes the masculine and the feminine rishons from each other.

*Charge*: masculine scent rishons have +5/9 electric charge, feminine-scent rishons have -4/9 electric charge.

*Colour*: both rishons can appear in three colours.

*Flavour*: the three combinations of two left- and one right-handed (total charge +2/3) as well as the three combinations of two right- and one left-handed (total charge -1/3) bound rishons compose the six flavours.

*Spin:* rishons are fermions. An additional one quarter spin is attributed to the clips, that bind rishons to each other.

*Quarks*: bound rishon triplets. They are fermions, since the spins of the even number binding clips neutralise each other.

*Baryons*: fermions composed of three quarks.

*Mesons*: bosons composed of a quark and an antiquark.

*Leptons*: half-spin fermions, composed of an identical colour rishon and antirishon, with antiparallel spin. Their resulting half-spin is added from the spin of the binding clips. The three individual colours are assumed to form the three generations of leptons.

*Weak charged vector bosons*: integer spin bosons composed of opposite scent, identical colour rishon and antirishon pairs with parallel spin. The spins of the binding clips neutralise each other.

*Weak neutral vector boson*: integer spin boson composed of identical scent, identical colour rishon-antirishon pairs with parallel spin.

γ: a colourless and scentless combination of identical scent, identical colour rishon-antirishon pairs with parallel spin.

*Gluons*: integer spin bosons composed of identical scent, different colour, parallel spin rishon-antirishon pair triplet combinations.

*Discussion of the QSD model*

Like the former, and among them finally Harari's, theories neither this model can give an account of the masses of the particles and a series of other properties. The mass problem is especially acute in this model in the case of the mass difference between the electron and the neutrino, as well as between massive and massless particles of the same rishon (but different spin) composition.

As regards the colours, quarks can be made only of equal colour rishons, and both the leptons and bosons of complementary colour rishon-antirishon pairs (the model cannot explain the reasons). We have no explanation for the difference in the used colour sets between the colour of leptons in the three generations. Moreover, no clue for the mass difference between the particle generations, if they differ in their colour only. This is the most speculative element of the model. (Nevertheless, the mentioned previous model was based on no less speculations making the distinction between the three generations.)

In spite of the unexplained problems, the new model has a few advantages compared to the previous one. First of all, it provides a more complete and consistent picture of the full system of particles considered in the SM.

Our masucline and feminine scent rishons appear in three colours each, and their triplets form the flavours (instead of the colours in Harari's model). The previous rishon model could not give an account on the structure of the second and third generations of quarks and leptons, more precisely, on the quantum number that distinguished between the three generations. QSD model – although it operates with six fundamental particles (instead of two in that) – presumably can give.

To draw a consistent theory of fermions and bosons the new model had to introduce an additional one-fourth spin attributed to the binding of the rishons, a hypothetic, non-observable property just for the explanation.

The previous model constructed all the quarks and the leptons of Tohu and Vohu triplets. That model could not explain, why are leptons free, while quarks are confined. For this new model represents the leptons as rishon-antirishon pairs, it leaves open to find a reasonable explanation for this difference. The rishons in our QSD model are confined, similar to the quarks in QCD.

Another difference for the favour of the new, QSD model, that weak bosons are able to act between single rishon states (although rishons appear always confined). A remarkable feature of the new model, that it attributes a similar structure to the weak bosons and the gluons.

*References*

Darvas, G. (2005) Order, entropy and smmetry, *Symmetry: Culture and Science*, **16**, 1, 91-108.
Darvas, G. (2007) *Symmetry*, Basel-Boston-Berlin: Birkhäuser, xi+508 pp.
Darvas, G., Farkas, F. T. (2006) An artist's works through the eyes of a physicist - Graphic illustration of symmetries of particles, *Leonardo*, 39, 1, 51-57.
Harari, H. (1979) *Phys. Lett. B*, **86**, 83.